\documentclass[aps,prl,twocolumn,amsmath,amssymb]{revtex4}
\usepackage{graphicx}
\usepackage{amsmath}

\begin{document}

\title{$s$-wave Scattering Resonances Induced by Dipolar Interactions of Polar Molecules}
\author{Zhe-Yu Shi, Ran Qi and Hui Zhai}
\email{hzhai@mail.tsinghua.edu.cn}
\affiliation{Institute for Advanced Study, Tsinghua University, Beijing, 100084, China}
\date{\today}
\begin{abstract}

We show that $s$-wave scattering resonances induced by dipolar interactions in a polar molecular gas have a universal large and positive effective range, which is very different from Feshbach resonances realized in cold atoms before, where the effective range is either negligible or negative. Such a difference has important consequence in many-body physics. At high temperature regime,  a positive effective range gives rise to stronger repulsive interaction energy for positive scattering length, and weaker attractive interaction energy for negative scattering length. While at low-temperatures, we study polaron problem formed by single impurity molecule, and we find that the polaron binding energy increases at the BEC side and decreases at the BCS side. All these effects are in opposite to narrow Feshbach resonances where the effective range is negative.

\end{abstract}
\maketitle

Polar molecular gas with strong dipolar interaction is a system of great interest to both atomic and molecular physics and condensed matter physics communities. Recent development of stimulated Raman adiabatic passage (STIRAP)  technique has succeeded in creating a nearly degenerate gas of KRb molecules in its rovibrational ground state \cite{Jin}. Unfortunately, the KRb molecules is not stable against two-body decay into K$_2$ and Rb$_2$ molecules \cite{chemical}. Recently, theoretical calculation predicts that this problem can be avoided by choosing other combinations of alkali atoms, such as KCs, KNa, NaCs, NaRb and RbCs \cite{stable}. Applying STIRP technique to these species is now proceeded in many different laboratories all over the world. It is very promising that degenerate polar molecular gases can be realized within the next few years.

For these molecules their electric dipole moments can be polarized by an external electric field (say, along $\hat{z}$ direction). As the applied electric field increases, the dipole moment $d$ continuously increases from zero to the permanent dipole moment, and different molecules have different values of permanent dipole moment \cite{numbers}. Therefore the two-body interaction potential studied here can be modeled as
\begin{align}
V_{\text{D}}({\bf r})=\begin{cases} \frac{d^2(1-3\cos^2\theta)}{r^3} \, , & r>r_{\text{c}} \\ \infty \, , & r<r_{\text{c}}
\end{cases}
\end{align}
which contains a long range dipolar interaction and a short range non-universal potential. $\theta$ is the angle between $\hat{r}$ and $\hat{z}$. Here for simplicity, we use a hard core potential to mimic short-range behavior of very complicated realistic potential, where $r_{\text{c}}$ is the core size. We can introduce a length scale $D=md^2/(2\hbar^2)$, and then the typical dipole energy is given by $\hbar^2D/(2m\langle r\rangle^3)\approx (k_{\text{F}}D)E_{\text{F}}$. By controlling the electric field, one can tune $k_{\text{F}}D$ from zero to the regime $k_{\text{F}}D\approx 1$ \cite{numbers}, i.e. the dipole energy is comparable to the Fermi energy.

Understanding the two-body problem of two strongly interacting quantum dipoles is an important step toward understanding the rich many-body physics of this system.
Previously, the dipolar interaction effects in the high partial wave channel have been studied extensively, for instance, the $p$-wave pairing \cite{p-wave_Pairing,review} and $p$-wave and $d$-wave Fermi surface distortion \cite{p-wave_Distortion,review}. While the dipolar interaction effects in the $s$-wave channel have not been well studied, because they always vanish in simple mean-field treatment.

In fact, previous studies have shown that, as $D$ increases, each partial wave channel will display a series of resonances where the scattering length of each partial wave channel diverges, and we call them dipolar interaction induced resonances (DIIR) \cite{LiYou,DIR_Blume,DIR}. Among all these channels, resonances in the $s$-wave channel are the most pronounced. However, both the two-body properties and the many-body physics at DIIR have not been completely understood. In this letter we study both two-body properties of DIIR and two many-body problems of two-component fermionic molecules across a DIIR, that are, high-temperature regime with equal population, and single impurity in fully polarized Fermi gas. From these studies we want to emphasize the significant difference between $s$-wave DIIR and other $s$-wave resonances. Main conclusions are summarized as follows:

(i) Near $s$-wave resonances the effective range $r_0$ is large and positive. This is qualitatively different from a magnetically tuned Feshbach resonance (MFR) studied before, where $r_0$ is either negligible (wide MFR) or large but negative (narrow MFR). Across different DIIRs, $r_{0}/D$ is a universal function of $D/a_{\text{s}}$ and $r_{0}/D\approx 1.84$ at resonance. Therefore $k_{\text{F}}r_0$ is of the order of unity and the effect of $r_0$ can not be ignored in many-body physics.

(iii) At high temperature, the attractive interaction energy of a DIIR is significantly smaller for negative scattering length $a_{\text{s}}$ compared to a wide MFR; while for positive $a_{\text{s}}$, in the upper branch, the repulsive energy of DIIR is much larger. Both can be understood by analyzing energy dependence of the scattering length or the two-body energy levels.

(iii) At low temperature, we consider polaron formed by single impurity molecule immersed in the Fermi sea. We find that the polaron binding energy is larger compared to a wide MFR at the BEC side and the resonance regime, while it is smaller at the BCS side.

{\it Two-body Problem.} Using $D$ as length unit $r\rightarrow r/D$ and $E_{\text{D}}=\hbar^2/(mD^2)$ as energy unit $E\rightarrow E/E_{\text{D}}$, the two-body Schr\"odinger equation can be written in dimensionless form
\begin{equation}
\left[-\frac{1}{2}\nabla^2+\frac{1-3\cos^2\theta}{r^3}\right]\Psi=E\Psi \label{schrodinger}
\end{equation}
in the regime $r>r_{\text{c}}$ and the boundary condition is given by $\Psi(r=r_{\text{c}})=0$. Expanding the wave function in terms of spherical harmonics as
$\Psi(r,\theta,\varphi)=\sum_{lm}\frac{1}{r}R_{lm}(r)Y_{lm}(\theta,\varphi)$, the Schr\"odinger equation can be reduced to
\begin{equation}
-\frac{1}{2}\frac{d^2}{dr^2}R_{lm}+\frac{1}{r^2}H_{l,l^\prime}^{m}(r)R_{l^\prime m}(r)=E R_{lm}\label{radialE}
\end{equation}
where $H_{ll^\prime}^{(m)}=l(l+1)\delta_{ll^\prime}+\langle Y_{lm}|(1-3\cos^2\theta)|Y_{l^\prime m}\rangle/r$. $H_{ll^\prime}$ couples all $l$ to $l\pm 2$ and $m$ is still a good quantum number. Moreover, since the dipolar coupling decays ($1/r^3$) faster than the centrifugal barrier term ($1/r^2$), at large distance, coupling between different $l$ effectively vanishes. The asymptotic behavior of the scattering wave function still can be decoupled as different partial wave channels, so one can introduce a scattering length for each partial wave channel \cite{pseudopotential}.

Here we will focus on the $s$-wave channel since it has no centrifugal barrier and the interaction effect is expected to be the strongest. However, it is easy to note that $H_{00}=0$, which means one cannot obtain any interaction effect if directly projecting the Schr\"odinger equation Eq. (\ref{schrodinger}) into the $|l,m\rangle=|0,0\rangle$ state. The interaction effect in the $s$-wave channel comes from its coupling to higher partial wave channel. Following a simple second-order perturbation argument, one can obtain an effective potential from a virtual process between $|0,0\rangle$ and $|2,0\rangle$,
\begin{eqnarray}
-\frac{\langle 00|(1-3\cos^3\theta)|20\rangle \langle 20|(1-3\cos^2\theta)|00\rangle \frac{D^2}{r^6}}{\frac{6}{r^2}}\propto -\frac{D^2}{r^4}\nonumber.
\end{eqnarray}
Hence, as $D$ increases, the effective potential becomes deeper and deeper, which introduces a series of bound states and causes scattering resonances. This is the basic mechanism of DIIR. Since the energies of those intermediate states for this induced interaction are $\hbar^2l(l+1)/(m\langle r\rangle^2)\gg 2E_{\text{F}}$ for $l=2,4,6,\dots$, we expect that the presence of a Fermi sea and the Pauli blocking effect will not strongly affect the induced interaction, which justifies using the induce interaction in many-body studies.

Following Ref. \cite{DIR}, we first diagonalize the matrix $H_{ll^\prime}$ by a unitary matrix $X(r)$ as $X^\dag(r)H(r)X(r)=\Lambda(r)$. Translating the radial wave function as $\Phi_{l}=X^\dag_{ll^\prime}R_{l^\prime}$, Eq. (\ref{radialE}) will become
\begin{equation}
-\frac{1}{2}\frac{d^2}{dr^2}\Phi_{l}+q_{ll^\prime}(r)\frac{d}{dr}\Phi_{l^\prime}+v_{ll^\prime}(r)\Phi_{l^\prime}=E\Phi_{l}    \label{radialE2}
\end{equation}
where $q(r)=-\frac{1}{2}X^\dag(r) dX(r)/dr$ and $v(r)=\Lambda(r)/r^2-\frac{1}{2}X^\dag(r)d^2X(r)/dr^2$. At a large distance, because $H(r)$ becomes effectively diagonal, $X$ becomes an identity matrix and $\Phi_l$ still has the same asymptotic behavior as $R_{l}$. At an intermediate and short distance, the diagonal part of $v(r)$ contains the centrifugal potential contribution for all $l\neq 0$, which separates $s$-wave from other partial waves. Because of this energy separation, we can take the approximation that only keeps the diagonal terms of $q_{ll^\prime}$ and $v_{ll^\prime}$. The infinite number of coupled differential equation of Eq. (\ref{radialE}) is reduced to a set of decoupled equations
\begin{equation}
-\frac{1}{2}\frac{d^2}{dr^2}\Phi_{l}+q_{ll}(r)\frac{d}{dr}\Phi_{l}+v_{ll}(r)\Psi_{l}=E\Phi_{l} \label{El}
\end{equation}
In Eq. (\ref{El}), $v_{00}(r)\neq 0$ is the induced interaction in $s$-wave channel. Numerically solving Eq. (\ref{El}) for $l=0$, we obtain its asymptotic behavior of $\Phi_{l=0}$ as $\sin(kr+\delta_{k})$, where $k=\sqrt{2mE}/\hbar$.

\begin{figure}[tbp]
\includegraphics[height=2.4in, width=3.3 in]
{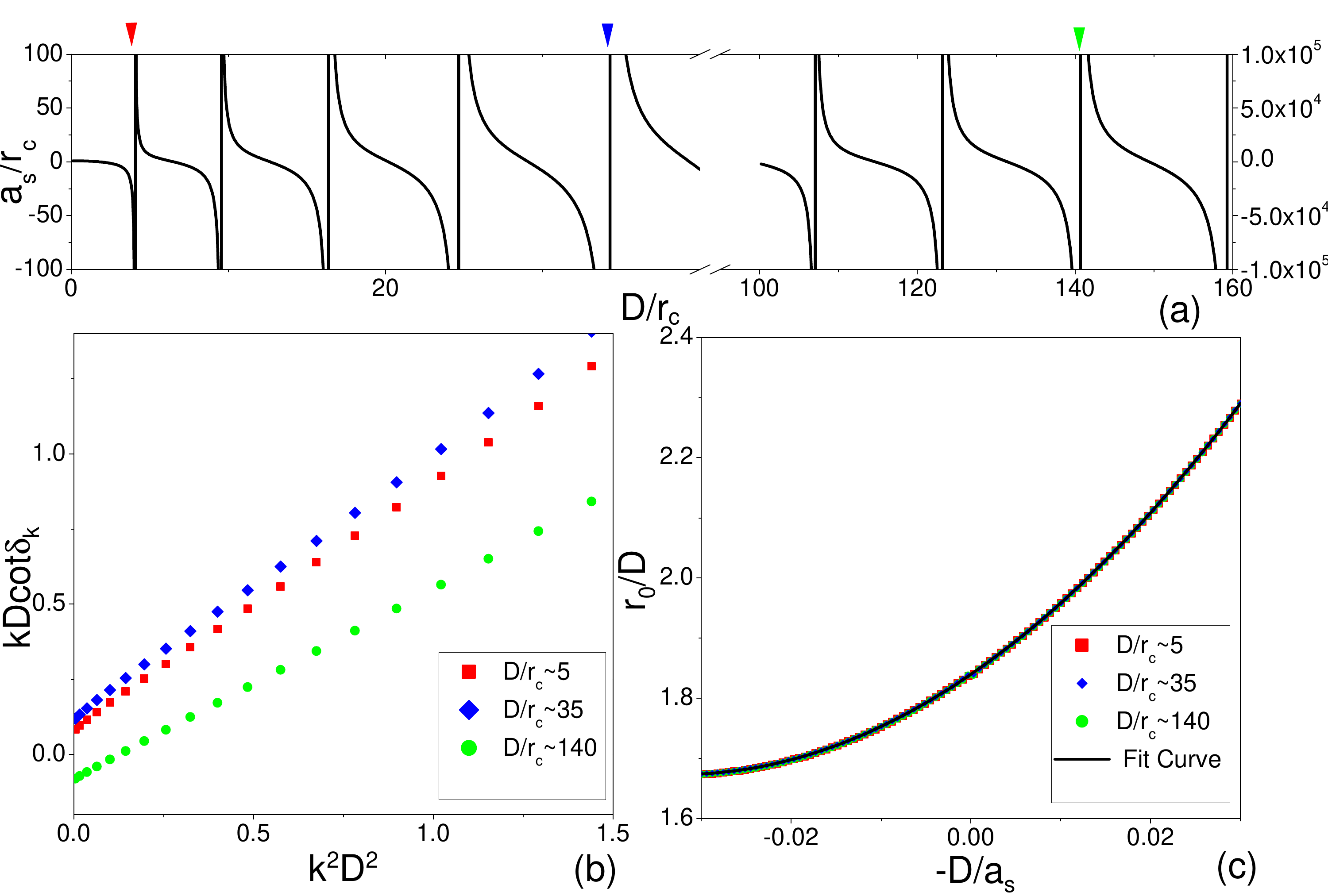}
\caption{(a) Scattering length $a_{\text{s}}/r_{\text{c}}$ as a function of $D/r_{\text{c}}$; (b) $kD\cot\delta_{k}$ as a function of $(kD)^2$ at different $D/r_{\text{c}}$ as marked in (a). (c) $r_0/D$ as a function of $-D/a_{\text{s}}$ across different resonances.  \label{two-body}}
\end{figure}

{\it Scattering Resonances and Effective Range.} Once we obtain $s$-wave phase shift from solving Eq. (\ref{El}), we can define $s$-wave scattering length as $a_{\text{s}}=-\lim_{k\rightarrow 0}\tan\delta_{k}/k$. We plot $a_{\text{s}}/r_{\text{c}}$ as a function of $D/r_{\text{c}}$ in Fig. (\ref{two-body})(a). It displays a series of resonances as $D/r_{\text{c}}$ increases, and the resonance window in $D/r_{\text{c}}$ axes gets wider for larger $D/r_{\text{c}}$. The locations of resonances can be determined by an intuitive WKB estimation \cite{DIR}. This approach precisely captures all $s$-wave resonances with relatively wide windows \cite{DIR_Blume}, and misses some extremely narrow ones which are due to a bound state in a higher even partial wave channel and coupled to $s$-wave channel by the ignored off-diagonal coupling in $q_{ll^\prime}$ and $v_{ll^\prime}$. We ignore them because they are too narrow to be studied experimentally.

We expand the phase shift as $k\cot\delta_{k}=-1/a_{\text{s}}+r_{0}k^2/2+\dots$. In Fig. \ref{two-body}(b) we find that $kD\cot\delta_{k}$ is a very good linear function of $(kD)^2$, and the positive slope means that $r_0$ is {\it positive}. In Fig. \ref{two-body}(c), we plot $r_0/D$ as a function of $D/a_{\text{s}}$ for different resonances with different $D/r_{\text{c}}$. It is very remarkable that all these curves across different resonances perfectly coincide with each other, which shows that $r_0/D$ is a universal function of $D/a_{\text{s}}$. In another word, the short-range physics ($r_{\text{c}}/D$) is irrelevant in determining $r_0$. Inspired by the relation between effective range and physical range of a square well potential, we fit the curve by $r_{0}/D=\alpha_0+\alpha_1 D/a_{\text{s}}+\alpha_2 (D/a_{\text{s}})^2$, and find $\alpha_0=1.8390\pm 0.0006$, and $\alpha_1=-10.168 \pm 0.002 $ and $\alpha_2=158.2\pm 0.1$. All the high order coefficients are several orders of magnitudes smaller. At resonance,  $r_{0}\approx 1.84D$. Noting that $D$ can be of the same order of $1/k_{\text{F}}$, we can therefore realize resonances in cold atom system with a large and positive effective range. This is very important of using cold atom to simulate nuclei and neutron matter \cite{INT,range}.

{\it High-Temperture Regime:} In the rest part of the paper, we will investigate the many-body effects of a positive effective range. We shall compare DIIR with a zero-range model for a wide MFR, which gives a constant phase shift $k\cot\delta_{k}=-1/a_{\text{s}}$ for all $k$, and $r_0=0$. First, we study the high-temperature regime by second order virial expansion, because resonant interaction manifests itself in interaction energy even above degenerate temperature, and can be measured by spectroscopy method easily.

\begin{figure}[tbp]
\includegraphics[height=1.4in, width=3.4 in]
{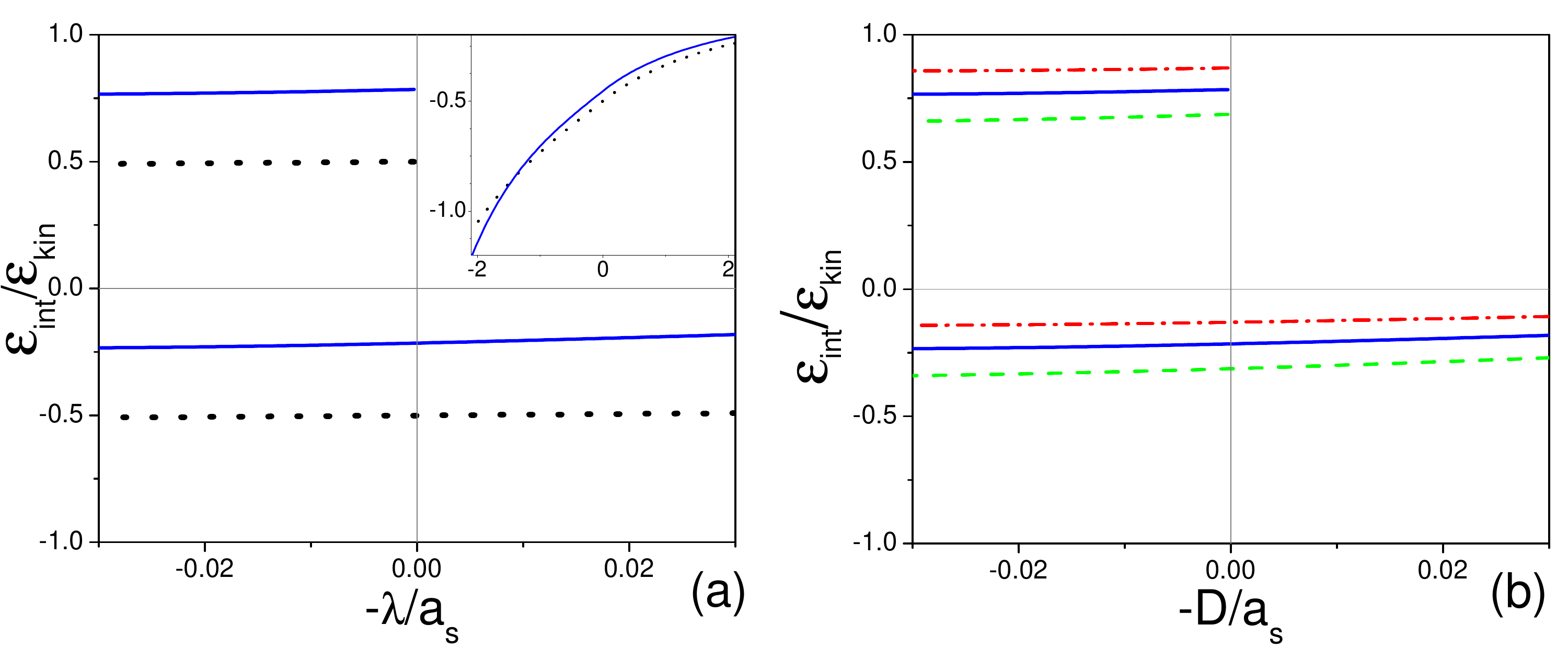}
\caption{(Color online) (a)  $\epsilon_{\text{int}}/\epsilon_{\text{kin}}$ as a function of $-\lambda/a_{\text{s}}$ for a DIIR (solid line) and a wide MFR (dotted line). $\lambda/D=1$. Inset: same plot for $\lambda/D=70$. (b)  $\epsilon_{\text{int}}/\epsilon_{\text{kin}}$ for DIIR with different temperatures. $\lambda/D=2$ for green blue dashed line; $\lambda/D=1$ for blue solid line and $\lambda/D=0.5$ for red dotted dashed line. Positive  $\epsilon_{\text{int}}/\epsilon_{\text{kin}}$ at positive $a_{\text{s}}$ side excludes the contribution from bound state. \label{Eint}}
\end{figure}

By second order virial expansion, we have
\begin{equation}
b_2=\sum e^{|E_{\text{b}}|/(k_{\text{b}}T)}+\int_{0}^{+\infty}\frac{dk}{\pi}\frac{d\delta_{k}}{dk}e^{-\lambda^2 k^2/(2\pi)} \label{b2}
\end{equation}
where we have ignored all the contributions from high partial waves, and $\lambda=\sqrt{2\pi\hbar^2/(mk_{\text{b}}T)}$. And the interaction energy is given by
\begin{equation}
\epsilon_{\text{int}}=\frac{3k_{\text{b}}Tn}{2}(n\lambda^3)\left[-\frac{b_2}{\sqrt{2}}+\frac{\sqrt{2}}{3}T\frac{\partial b_2}{\partial T}\right]
\end{equation}
In Fig. \ref{Eint}(a), we compare $\epsilon_{\text{int}}/\epsilon_{\text{kin}}$ for a DIIR with a wide MFR, and in Fig. \ref{Eint}(b), we show $\epsilon_{\text{int}}/\epsilon_{\text{kin}}$ for a DIIR with different temperatures, where $\epsilon_{\text{kin}}=3k_{\text{b}}Tn(n\lambda^3)/2$. In the side with $a_{\text{s}}>0$, negative $\epsilon_{\text{int}}$ includes the contribution from bound state, while positive $\epsilon_{\text{int}}$ excludes bound state contribution, which corresponds to the physics of "the upper branch".

First, for scattering states, it is known that, $\epsilon_{\text{int}}/\epsilon_{\text{kin}}$ approaches $\mp 0.5$ as one approaches a wide MFR from negative and positive side of $a_{\text{s}}$ \cite{Erich}. While for a DIIR, we find for $a_{\text{s}}<0$, the attractive interaction energy is weaker for a DIIR, and for $a_{\text{s}}>0$, the repulsive interaction energy is stronger. This is strongly in contrast to a narrow MFR ($r_0<0$) where the opposite effect is found by Ref. \cite{Ho}. This can be understood from an argument based on energy dependence on scattering length. Let us consider a situation that $a_{\text{s}}$ not very close to resonance, and $r_0$ is non-zero, but is still small compared to $\lambda^2/a_{\text{s}}$. Because $a(\bar{k})=1/(1/a_{\text{s}}-r_0 \bar{k}^2/2)$, ($\bar{k}$ denotes a thermal average of $k$, and is of the order of $1/\lambda$), for $a_{\text{s}}<0$, one finds $|a(\bar{k})|<|a_{\text{s}}|$ for $r_0>0$, and $|a(\bar{k})|>|a_{\text{s}}|$ for $r_0<0$; And for $a_{\text{s}}>0$, one finds $|a(\bar{k})|>|a_{\text{s}}|$ for $r_0>0$, and $|a(\bar{k})|<|a_{\text{s}}|$ for $r_0<0$. This is consistent with $r_0$ dependence of interaction energy shown in Fig. \ref{Eint}(a).

An alternative way to understand this result is through analyzing energy levels. Let us consider two-atom confined in a large hard sphere so that the two-body wave function $\psi(r)$ has to vanish at $r=R$. The energy levels of $s$-wave states is determined by $E_n=\hbar^2 k_n^2/m$, where $k_n$ satisfies the equation $k_nR+\delta_{k_n}=n\pi$. In the limit $a_{\text{s}}\rightarrow 0^{-}$, $\delta_k=0$ and $k^{0^-}_n=n \pi/R$; while for $a_{\text{s}}\rightarrow 0^{+}$, $\delta_{k}=\pi$ and $k^{0^{+}}_n=(n-1)\pi/R$. For a wide MFR, at resonance, $\delta_k=\pi/2$, and all $k_n$ is given by $(n-1/2)\pi/R$, which is responsible for $\epsilon_{\text{int}}/\epsilon_{\text{kin}}=\pm 0.5$ at resonance \cite{Ho}. While for a DIIR, at resonance, $k$ satisfies $kR+\arctan(2/(r_{0}k))=n\pi$. At a given temperature, the system is dominated by states with $k\sim 1/\lambda$. Hence, $\arctan(2/(r_{0} k))$ is always smaller than $\pi/2$ if $r_0>0$ and it decreases to zero as temperature increases. Therefore, the downshifts of the energy levels $E^{0^{-}}_n-E_n$ are smaller compared to a wide MFR, and they decrease as temperature increases. And the upshifts of the energy levels $E_n-E^{0^+}_n$ are larger compared to a wide MFR, and they increase as temperature increases. This is also consistent with temperature dependence of $\epsilon_{\text{int}}/\epsilon_{\text{kin}}$ as shown in Fig. \ref{Eint}(b).

The effective range $r_0$ also has effect on the bound state energy. The bound state energy is given by the pole of scattering amplitude $f(E)=1/(1/a_{\text{s}}-r_0 E/2-i\sqrt{E})$, which gives
\begin{equation}
E=-\frac{\hbar^2}{ma^2_{\text{s}}}\frac{1-\sqrt{1-2r_0/a_{\text{s}}}}{r_0/a_{\text{s}}}\simeq -\frac{\hbar^2}{ma^2_{\text{s}}} \left(1+\frac{r_0}{2a_{\text{s}}}\right).
\end{equation}
For the same $a_{\text{s}}$, the binding energy $|E|$ is larger with positive $r_0$. For same reason, for a narrow MFR with negative $r_0$, the binding energy is smaller. In the inset of Fig. \ref{Eint}(a) we show that at a lower temperature, the attractive interaction energy of a DIIR will finally exceed that of a wide MFR. This is because for a lower temperature, the contribution of the lowest bound state will gradually become dominative, and the larger binding energy finally  overwhelms the smaller contribution from the scattering states.

\begin{figure}[tbp]
\includegraphics[height=1.4in, width=3.4 in]
{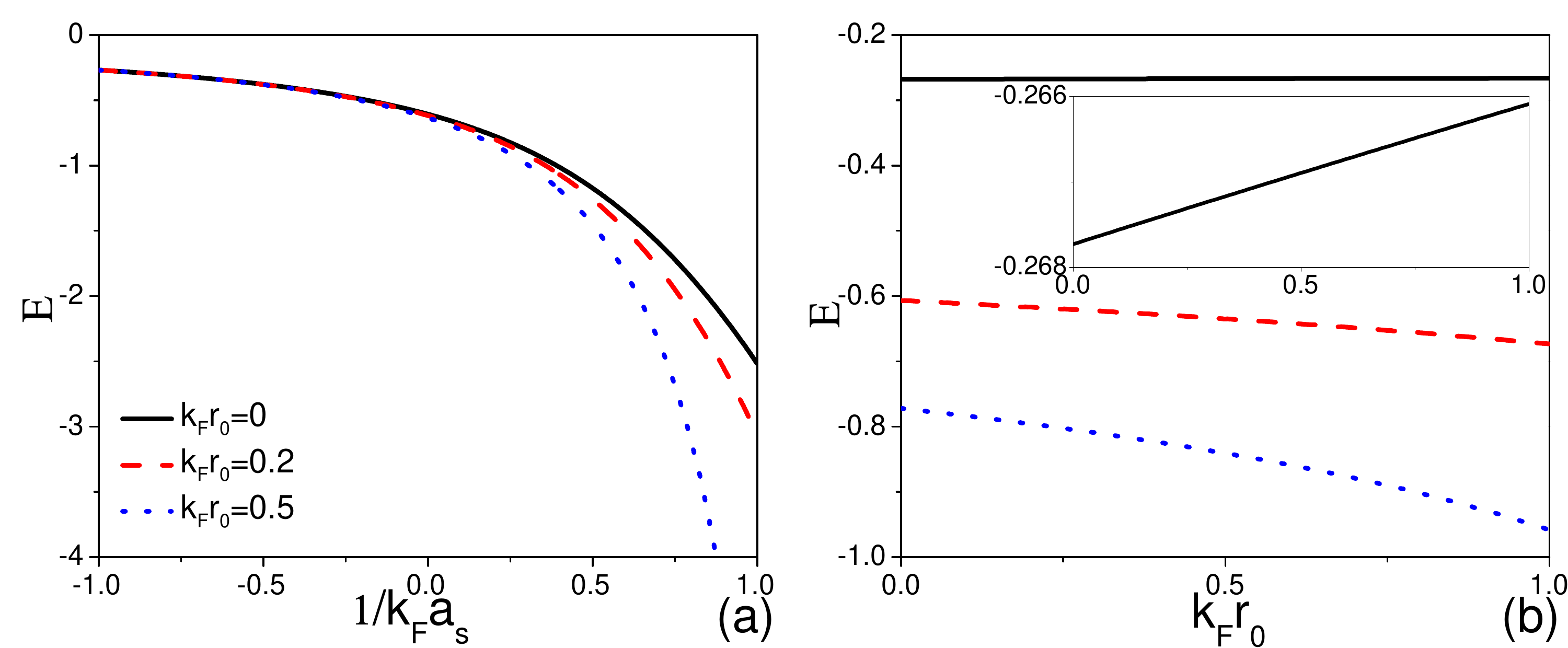}
\caption{(Color online) (a) Polaron energy $\mathcal{E}/E_{\text{F}}$ as a function of $-1/(k_{\text{F}}a_{\text{s}})$ for $k_\text{F} r_0=0$ (wide resonance, solid line); $k_\text{F} r_0=0.2$ (dashed line) and $k_\text{F} r_0=0.5$ (dotted line); (b)  $\mathcal{E}/E_{\text{F}}$ as a function of $k_F r_0$ for $1/(k_{\text{F}}a_{\text{s}})=-1$ (solid line); $1/(k_{\text{F}}a_{\text{s}})=0$ (dashed line) and $1/(k_{\text{F}}a_{\text{s}})=0.2$ (dotted line) \label{polaron}}
\end{figure}

{\it Single Impurity Problem at Low Temperature:} In general, the low-temperature many-body physics at resonance is too sophisticated to be studied by simple method. Here, in order to illustrate that the effect of $r_0$ at low temperature could be different from that in high temperature, we consider a simpler situation where one single impurity molecule is dressed by a Fermi sea of majority molecules and form a polaron. For wide resonance, such a problem can be well described by a variational wave function or $T$-matrix approach \cite{polaron}. Here we apply similar approach to DIIR and obtain following self-consistent equation for polaron energy $\mathcal{E}$
\begin{eqnarray}
\mathcal{E}&=&\sum_{|{\bf q}|<|{\bf k_{\text{F}}}|}\frac{1}{\frac{Vm}{4\pi}\left[\frac{1}{a_{\text{s}}}-\frac{mr_{0}}{2}(\mathcal{E}+\epsilon^{\uparrow}_{\mathbf{q}}-\epsilon^{\text{b}}_{\mathbf{q}})\right]
+I(\mathcal{E})}
\end{eqnarray}
where $I(\mathcal{E},{\bf q})=\sum_{|{\bf k}|>k_{\text{F}}}\left(\frac{1}{\epsilon^{\uparrow}_{\mathbf{k}}+\epsilon^{\downarrow}_{\mathbf{p}+\mathbf{q}-\mathbf{k}}
-\epsilon^{\uparrow}_{\mathbf{q}}-\epsilon^{\downarrow}_{\mathbf{p}}-\mathcal{E}} -\frac{1}{\epsilon^\text{r}_{\mathbf{k}}}\right)-\sum_{|{\bf k}|<k_\text{F}}\frac{1}{\epsilon^\text{r}_{\mathbf{k}}}$, and $\epsilon^\text{r}_{\mathbf{k}}={\bf k}^2/m$, $\epsilon^{\text{b}}_{\mathbf{k}}={\bf k}^2/(4m)$. The details for obtaining this equation is similar as the same impurity problem across a narrow MFR \cite{Qi}. We find that the effective range has more dramatic effect on polaron binding energy at the BEC side than at the BCS side, as shown in Fig. \ref{polaron}(a). And we find that at the BCS side, a positive $r_0$ will decrease the polaron binding energy $|\mathcal{E}|$, while it will increase $|\mathcal{E}|$ at the resonance regime and the BEC side. Such an effect is also opposite to polaron nearby a narrow MFR \cite{Qi}, and for narrow resonance, such effect has been observed in a recent experiment \cite{Grimm}. In both cases, it is caused by and can be understood from the energy dependence on scattering length \cite{Qi}. It will be very interesting to observe the counter effect in dipolar gases of polar molecules, so that the effects of effective range in many-body systems will be established in a comprehensive way.

{\it Acknowledgements.} We thank Zeng-Qiang Yu and Xiao-Ling Cui for very helpful discussions, and we thank Tin-Lun Ho for careful reading the manuscript. This work is supported by Tsinghua University Initiative Scientific Research Program, NSFC under Grant No. 11004118 (HZ), No. 11174176 (HZ), No. 11104157 (RQ) and NKBRSFC under Grant No. 2011CB921500.

\end{document}